\documentclass[prd,twocolumn,amsmath,amssymb]{revtex4}
\usepackage{graphicx}
\begin{document}

\title{
\boldmath Direct Measurements of the Branching Fractions for
Inclusive $K^\pm$ and Inclusive Semileptonic Decays of $D^+$ and
$D^0$ Mesons}
\author{
\vspace{0.35cm}
 {\large BES Collaboration}\\
\vspace{0.55cm} \normalsize M.~Ablikim$^{1}$, J.~Z.~Bai$^{1}$,
Y.~Ban$^{12}$, X.~Cai$^{1}$, H.~F.~Chen$^{16}$, H.~S.~Chen$^{1}$,
H.~X.~Chen$^{1}$, J.~C.~Chen$^{1}$, Jin~Chen$^{1}$,
Y.~B.~Chen$^{1}$
% ...
\\
{ \vspace{0.25cm} \it
$^{1}$ Institute of High Energy Physics, Beijing 100049, People's Republic of China\\
% ...
$^{12}$ Peking University, Beijing 100871, People's Republic of China\\
% ...
$^{16}$ University of Science and Technology of China, Hefei 230026, People's Republic of China\\
% ...
}}

\vspace{0.5cm}
\begin{abstract}
With singly-tagged $\bar D$ samples selected from the data collected
at and around 3.773 GeV with the BESII detector at the BEPC
collider, we have measured the branching fractions for the inclusive
$K^\pm$ decays of $D^+$ and $D^0$ mesons, which are $BF(D^+\to K^-X)
= (24.7
 \pm 1.3 \pm 1.2)\%$, $BF(D^+\to K^+X) = (6.1 \pm 0.9 \pm 0.4) \%$,
$BF(D^0\to K^-X) = (57.8 \pm 1.6 \pm 3.2) \%$
 and
$BF(D^0\to K^+X) = (3.5 \pm 0.7 \pm 0.3) \%$, respectively.
 We have also measured the branching fractions for the inclusive semileptonic decays of $D^+$ and $D^0$ mesons to
be $BF(D^+ \to e^+ X)=(15.2 \pm 0.9 \pm 0.8)\%$ and $BF(D^0 \to e^+
X) =(6.3 \pm 0.7 \pm 0.4) \%$. These yield the ratio of their
partial widths to be $\Gamma(D^+ \to e^+X)/\Gamma(D^0 \to e^+X)=0.95
\pm
 0.12 \pm 0.07$.

\end{abstract}

\maketitle
 \oddsidemargin
-0.2cm \evensidemargin -0.2cm

\newpage
\section{\bf Introduction}

Measurements of the branching fractions for inclusive $K^\pm$ decays
of charged and neutral $D$ mesons give us some useful information
about the relative strengths of the inclusive decay processes.
Measurements of the branching fractions for inclusive semileptonic
decays of $D^+$ and $D^0$ mesons provide some tests of the
contributions from different decay diagrams and provide some
information for understanding the origin of the lifetime difference
of $D^+$ and $D^0$ mesons. These measurements can serve as a check
on the sum of the measured branching fractions \cite{pdg} for the
exclusive decay modes of $D$ mesons, which can guide one to search
for some new decay modes. Moreover, these measurements can also
provide helpful information for the studies of the $B$ meson decays.

In this Letter, we report direct measurements of the branching
fractions for the inclusive decays $D^+\to K^-X$~($X$ = any
particles) and $D^0\to K^-X$, as well as the inclusive decays
$D^+\to K^+X$ and $D^0\to K^+X$. We also report direct measurements
of the branching fractions for the inclusive semileptonic decays
$D^+\to e^+X$ and $D^0 \to e^+ X$. Then, we determine the ratio of
their partial widths, $\Gamma(D^+ \to e^+X)/\Gamma(D^0 \to e^+X)$.
These measurements are made by analyzing the data of about 33
pb$^{-1}$ collected at and around 3.773 GeV with the BESII detector
at the BEPC collider.

\section{BESII detector}
The BESII is a conventional cylindrical magnetic detector
\cite{bes2} operated at the Beijing Electron-Positron Collider
(BEPC) \cite{bepc}. A 12-layer Vertex Chamber (VC) surrounding the
beryllium beam pipe provides input to the event trigger, as well as
coordinate information. A forty-layer main drift chamber (MDC)
located just outside the VC yields precise measurements of charged
particle trajectories with a solid angle coverage of $85\%$ of
4$\pi$; it also provides ionization energy loss ($dE/dx$)
measurements which are used for particle identification. Momentum
resolution of $1.7\%\sqrt{1+p^2}$ ($p$ in GeV/$c$) and $dE/dx$
resolution of $8.5\%$ for Bhabha scattering electrons are obtained
for the data taken at $\sqrt{s}=3.773$ GeV. An array of 48
scintillation counters surrounding the MDC measures the time of
flight (TOF) of charged particles with a resolution of about 180 ps
for electrons. Outside the TOF, a 12 radiation length, lead-gas
barrel shower counter (BSC), operating in limited streamer mode,
measures the energies of electrons and photons over $80\%$ of the
total solid angle with an energy resolution of
$\sigma_E/E=0.22/\sqrt{E}$ ($E$ in GeV) and spatial resolutions of
$\sigma_{\phi}=7.9$ mrad and $\sigma_Z=2.3$ cm for electrons. A
solenoidal magnet outside the BSC provides a 0.4 T magnetic field in
the central tracking region of the detector. Three double-layer muon
counters instrument the magnet flux return and serve to identify
muons with momentum greater than 500 MeV/$c$. They cover $68\%$ of
the total solid angle.

\section{\bf Data analysis}
The $\psi(3770)$ resonance is produced in electron-positron
($e^+e^-$) annihilation at the center-of-mass energy of about 3.773
GeV. It decays to $D\bar D$ pairs ($D^0\bar D^0$ and $D^+D^-$) with
a large branching fraction \cite{bf_psipp_to_dd,cleo_dd}. If one $D$
meson is fully reconstructed~(called singly-tagged $\bar D$ event),
there must be another $D$ meson in the system recoiling against the
singly-tagged $\bar D$ meson. This makes it possible to directly
measure the branching fractions for the inclusive decays of $D^+$
and $D^0$ mesons. To do so, we first reconstruct the singly-tagged
$\bar D$ mesons \cite{kpiev,dcrs,k0ev,kstev} from the data. In the
system recoiling against the singly-tagged $\bar D$ mesons, we
select the candidates for the inclusive decays of $D^+$ and $D^0$
mesons, and measure the branching fractions for these decays
directly. Throughout the Letter, charge conjugates are implied.

\subsection{\bf Event selection}

All charged tracks used in data analysis are required to be well
reconstructed in the MDC with good helix fits, and to satisfy
$|$cos$\theta|<0.85$, where $\theta$ is the polar angle of the
charged track. Each track, except for those from $K^0_S$, must
originate from the interaction region, which requires that the
closest approach to the interaction point of a charged track in the
$xy$-plane is less than 2.0 cm and in the $z$ direction is less than
20.0 cm.

The charged kaons, pions and electrons are identified with the
$dE/dx$, TOF and BSC measurements. With these measurements, we
calculate the confidence levels ($CL_K$, $CL_{\pi}$ and $CL_e$) for
a kaon, a pion and an electron hypotheses. The pion candidate is
required to have $CL_{\pi}>0.001$. The kaon candidate is required to
satisfy $CL_K>CL_{\pi}$ and $CL_{K}>0.001$. The electron candidate
is required to satisfy $CL_e>0.01$ and
$CL_e/(CL_e+CL_K+CL_{\pi})>0.8$.

The neutral kaons are reconstructed through the decay
$K^0_S\to\pi^+\pi^-$. The candidate tracks used for the $K^0_S$
reconstruction are required to originate from the interaction region
$V_{xy}<8.0$ cm and $|V_z|<20.0$ cm, where $V_{xy}$ and $|V_z|$ are
the closet approaches of the charged track in the $xy$-plane and $z$
direction.

The neutral pions are reconstructed through the decay $\pi^0 \to
\gamma \gamma$. In order to reduce background, we require that the
good photon candidates satisfy the following criteria: the energy
deposited in the BSC of each photon is greater than 70 MeV, the
electromagnetic shower starts in the first five readout layers, the
angle between the photon and the nearest charged track is greater
than $22^\circ$ \cite{kpiev,k0ev,kstev}, the angle between the
cluster development direction and the photon emission direction is
less than $37^\circ$ \cite{kpiev,k0ev,kstev}.

\subsection{ \bf Singly-tagged $\bar{D}$ samples}
The singly-tagged $D^-$ mesons are reconstructed in nine hadronic
decays of $K^+\pi^-\pi^-$, $K^0\pi^-$, $K^0K^-$, $K^+K^-\pi^-$,
$K^0\pi^-\pi^-\pi^+$, $K^0\pi^-\pi^0$, $K^+\pi^-\pi^-\pi^0$,
$K^+\pi^-\pi^- \pi^-\pi^+$ and $\pi^-\pi^-\pi^+$. The singly-tagged
$\bar{D}^0$ mesons are reconstructed in three hadronic decays of
$K^+\pi^-$, $K^+\pi^-\pi^0$ and $K^+\pi^-\pi^-\pi^+$. Total $5321
 \pm 149{\rm (stat.)} \pm 160{\rm (sys.)}$ singly-tagged $D^-$ mesons
and $7033 ~\pm~ 193{\rm (stat.)} ~\pm~ 316{\rm (sys.)}$
singly-tagged $\bar D^0$ mesons are reconstructed from the data by
examining the invariant masses of the $mKn\pi(m=0,1,2;~n=0,1,2,3,4)$
combinations. Refs.~\cite{kpiev,dcrs,k0ev,kstev} described the
details about the selection of the singly-tagged $D^-$ and
$\bar{D}^0$ mesons.

\subsection{\bf Candidates for $D^+\to e^+X$ and $D^0\to e^+X$}

The electron candidates are selected from the surviving charged
tracks in the system recoiling against the singly-tagged $\bar D$
mesons. The selected candidate tracks are required to originate from
the same vertex as those decay from the singly-tagged $\bar D$
mesons by requiring $\delta z < 2\sigma_{\delta z}$ (2.0 cm), where
$\delta z$ is the minimum distance in the $z$ direction between the
candidate track and those decay from the singly-tagged $\bar{D}$
mesons, $\sigma_{\delta z}$ is the standard deviation of the $\delta
z$ distribution.

Electrons with the charge opposite to the charm of the singly-tagged
$\bar D$ meson are defined as the right-sign electrons; on the
contrary, they are defined as the wrong-sign electrons. The
wrong-sign electrons accounting for the decay $\pi^0\to\gamma
e^+e^-$ and $\gamma$ conversions, etc. are used to estimate the
charge-symmetric background in the selected right-sign electrons.

\begin{figure}[htbp]
\begin{center}
  \includegraphics[width=8cm,height=7cm]
{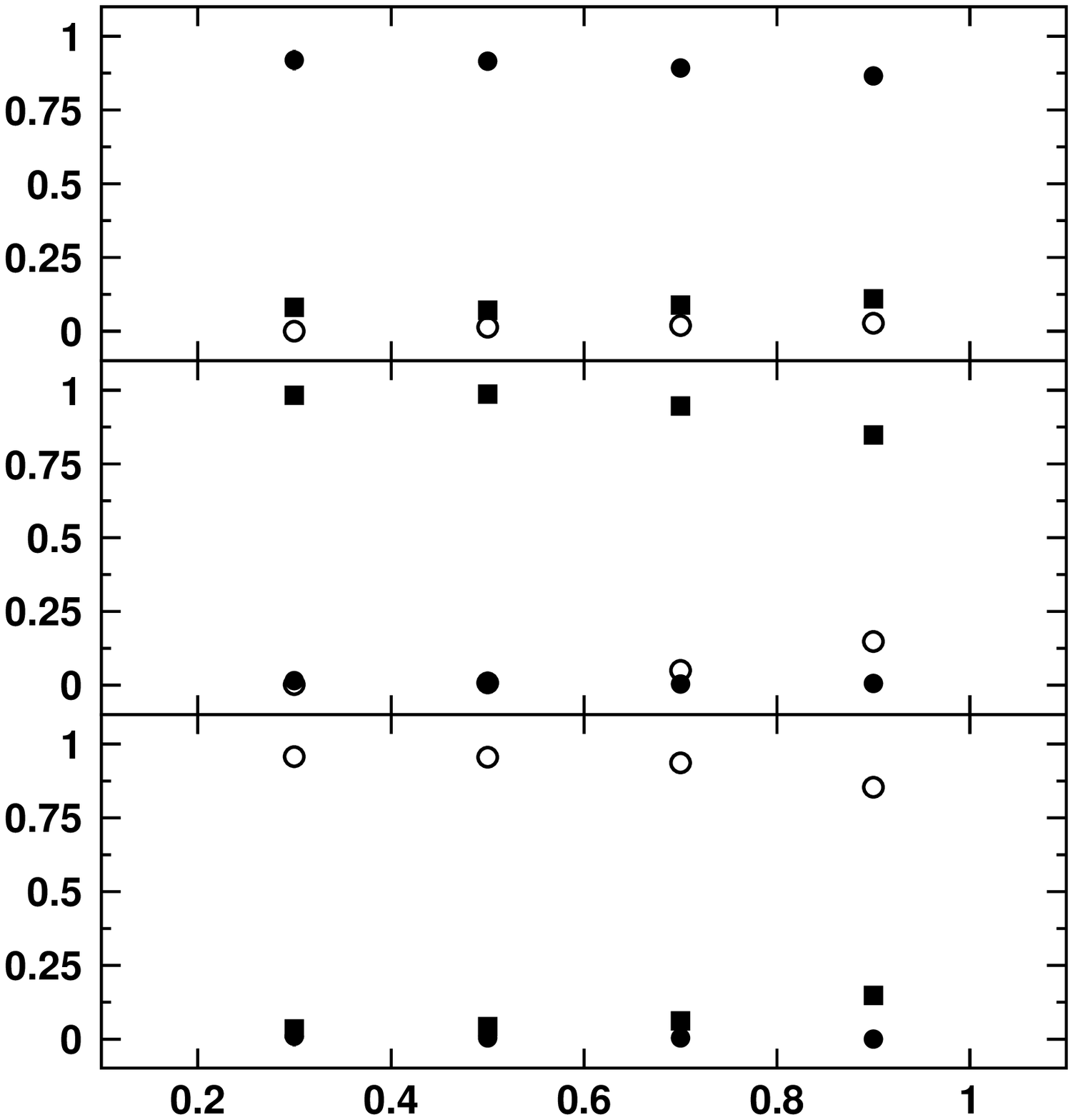}
      \put(-185,170){\small \bf (a)}
      \put(-185,115){\small \bf (b)}
      \put(-185,62){\small \bf (c)}
      \put(-150,0){\bf Momentum (GeV/$c$)}
      \put(-230,100){\normalsize \rotatebox{90}{$\epsilon_{a,i}$ or $f_{b \to a,i}$}}
\caption {\small The efficiencies $\epsilon_{a,i}$ and misidentified
probabilities $f_{b \to a,i}$ (see text).}
\label{fig:eff}
\end{center}
\end{figure}

The electron, kaon and pion may be misidentified to each other with
different probabilities in different momentum ranges. In order to
remove such background from the electron samples, we need to know
the yield $N_{\pi,i}^{\rm obs}$ (throughout the Letter, the $i$
denotes the $i$th momentum range) of pions and the yield
$N_{K,i}^{\rm obs}$ of kaons besides the yield $N_{e,i}^{\rm obs}$
of electrons. They are also selected in the system recoiling against
the singly-tagged $\bar D$ mesons, and similarly separated into the
right-sign and wrong-sign pions or kaons. In addition, we also need
to know the efficiencies $\epsilon_{e,i}$, $\epsilon_{\pi,i}$ and
$\epsilon_{K,i}$ of identifying electron, pion and kaon, as well as
the probability $f_{b \to a,i}$ of misidentifying the particle $b$
as $a$. Here, $a$ and $b=e$, $\pi$, $K$, but $b\ne a$. These
efficiencies and misidentified probabilities in each momentum range
are shown in Fig. \ref{fig:eff}. In Fig. \ref{fig:eff}(a), the dots
with error bars show the efficiencies of identifying electron, and
the squares (circles) with error bars show the probabilities of
misidentifying electron as pion (kaon). In Fig. \ref{fig:eff}(b),
the squares with error bars show the efficiencies of identifying
pion, and the dots (circles) with error bars show the probabilities
of misidentifying pion as electron (kaon). In Fig. \ref{fig:eff}(c),
the circles with error bars show the efficiencies of identifying
kaon, and the dots (squares) with error bars show the probabilities
of misidentifying kaon as electron (pion). The efficiencies and
misidentified probabilities are estimated by analyzing pure
electron, pion and kaon samples. The electron sample is selected
from the radiative Bhabha events. The pion and kaon samples are
selected from the $J/\psi\to\omega\pi^+\pi^-$ and $J/\psi\to\phi
K^+K^-$ events, respectively.

\begin{figure}[htbp]
\begin{center}
  \includegraphics[width=8cm,height=6cm]
{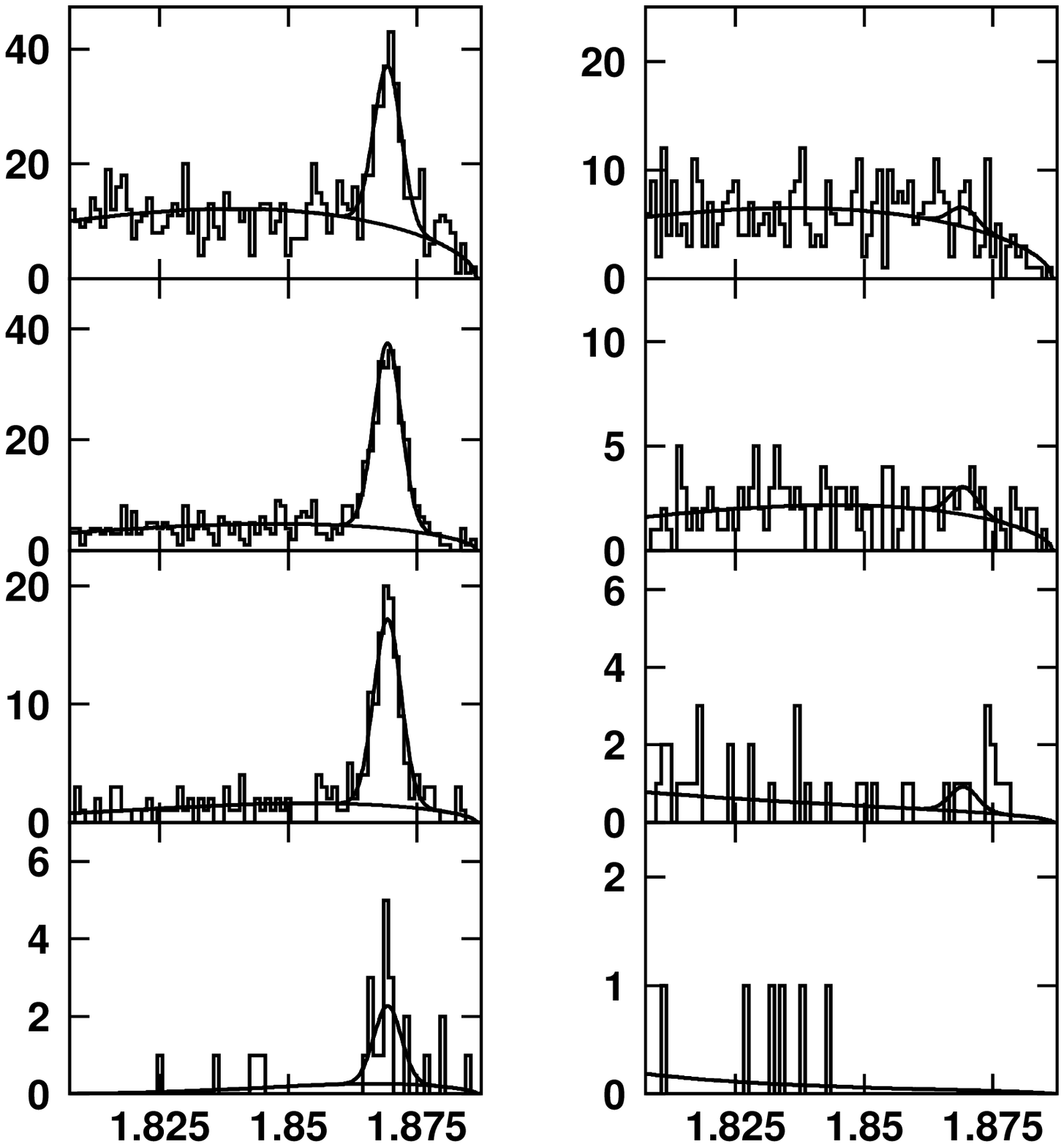}
      \put(-183,151){\small \bf a}
      \put(-183,114){\small \bf b}
      \put(-183,80){\small \bf c}
      \put(-183,43){\small \bf d}
      \put(-80,151){\small \bf a$^\prime$}
      \put(-80,114){\small \bf b$^\prime$}
      \put(-80,80){\small \bf c$^\prime$}
      \put(-80,43){\small \bf d$^\prime$}
      \put(-170,0){\bf \small Invariant Mass (GeV/$c^2$)}
      \put(-220,50){\rotatebox{90}{\bf \small Events/(0.001GeV/$c^2$)}}
\caption {\small The distributions of the invariant masses of the
$mKn\pi$ combinations for the events in which the candidates for the
right-sign (left) and wrong-sign (right) electrons with the momentum
within (a) 0.2-0.4 GeV/$c$, (b) 0.4-0.6 GeV/$c$, (c) 0.6-0.8 GeV/$c$
and (d) 0.8-1.0 GeV/$c$ ranges are observed in the system recoiling
against the $mKn\pi$ combinations (for the singly-tagged $ D^-$
mesons). } \label{fig:dpex}
\end{center}
\end{figure}

\begin{figure}[htbp]
\begin{center}
  \includegraphics[width=8cm,height=6cm]
{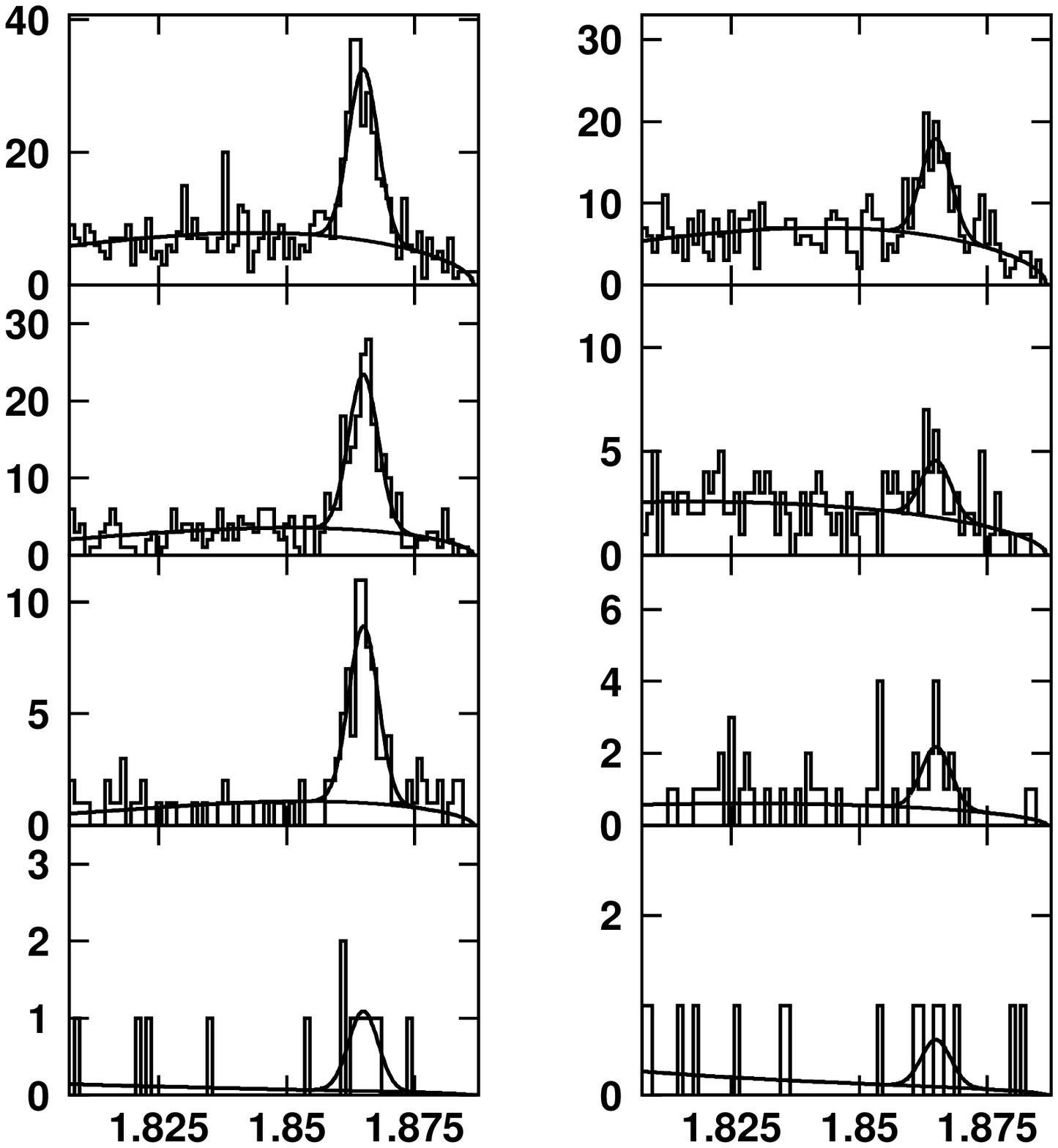}
       \put(-183,151){\small \bf a}
      \put(-183,114){\small \bf b}
      \put(-183,80){\small \bf c}
      \put(-183,43){\small \bf d}
      \put(-80,151){\small \bf a$^\prime$}
      \put(-80,114){\small \bf b$^\prime$}
      \put(-80,80){\small \bf c$^\prime$}
      \put(-80,43){\small \bf d$^\prime$}
      \put(-170,0){\bf \small Invariant Mass (GeV/$c^2$)}
      \put(-220,50){\rotatebox{90}{\bf \small Events/(0.001GeV/$c^2$)}}
 \caption {\small The distributions of the invariant masses of the $mKn\pi$
combinations for the events in which the candidates for the
right-sign (left) and wrong-sign (right) electrons with the momentum
within (a) 0.2-0.4 GeV/$c$, (b) 0.4-0.6 GeV/$c$, (c) 0.6-0.8 GeV/$c$
and (d) 0.8-1.0 GeV/$c$ ranges are observed in the system recoiling
against the $mKn\pi$ combinations (for the singly-tagged $\bar D^0$
mesons). } \label{fig:d0ex}
\end{center}
\end{figure}

\begin{figure}[htbp]
\begin{center}
  \includegraphics[width=8cm,height=6cm]
{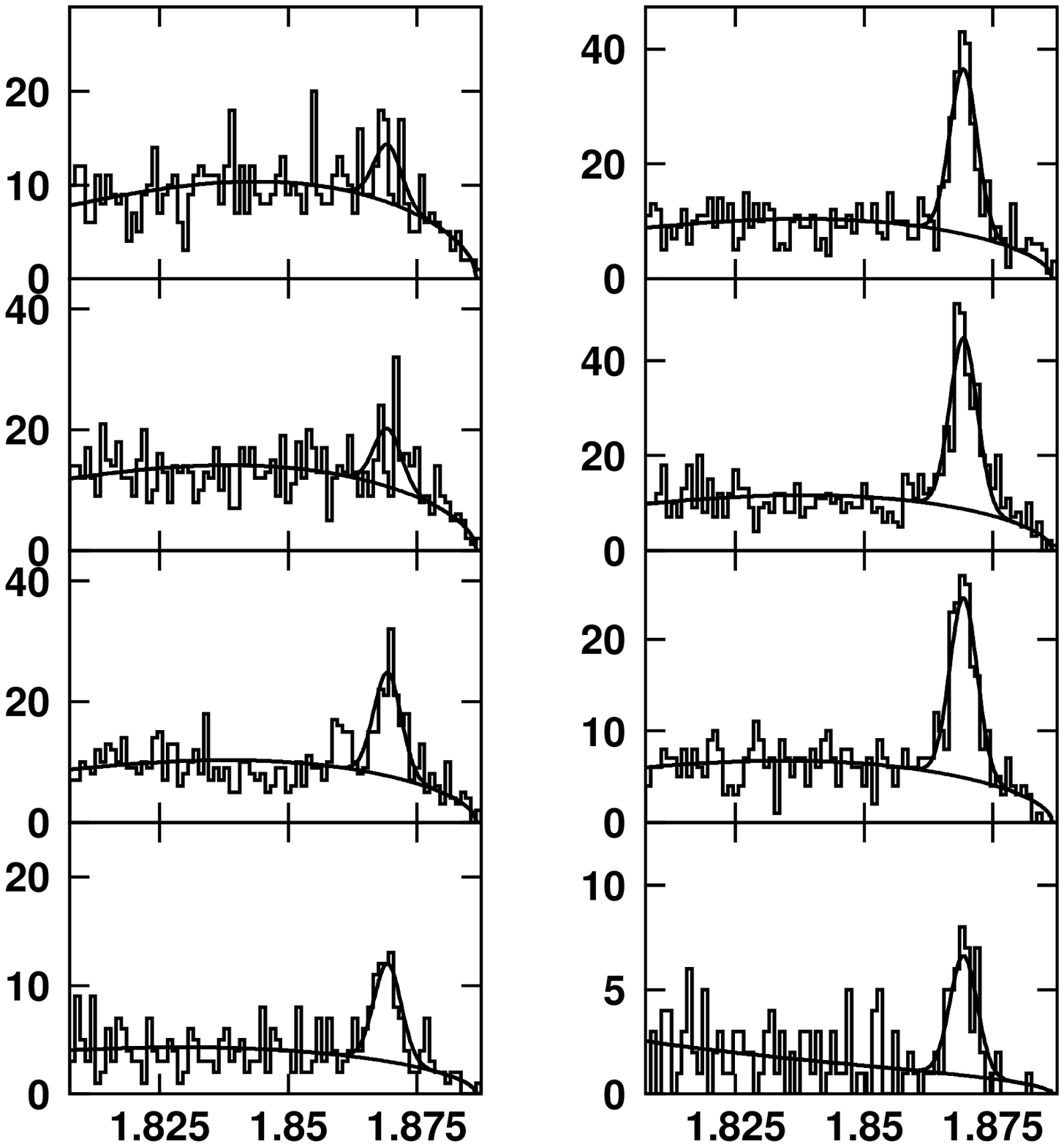}
       \put(-183,151){\small \bf a}
      \put(-183,114){\small \bf b}
      \put(-183,80){\small \bf c}
      \put(-183,43){\small \bf d}
      \put(-80,151){\small \bf a$^\prime$}
      \put(-80,114){\small \bf b$^\prime$}
      \put(-80,80){\small \bf c$^\prime$}
      \put(-80,43){\small \bf d$^\prime$}
      \put(-170,0){\bf \small Invariant Mass (GeV/$c^2$)}
      \put(-220,50){\rotatebox{90}{\bf \small Events/(0.001GeV/$c^2$)}}
\caption {\small The distributions of the invariant masses of the
$mKn\pi$ combinations for the events in which the candidates for the
right-sign (left) and wrong-sign (right) kaons with the momentum
within (a) 0.2-0.4 GeV/$c$, (b) 0.4-0.6 GeV/$c$, (c) 0.6-0.8 GeV/$c$
and (d) 0.8-1.0 GeV/$c$ ranges are observed in the system recoiling
against the $mKn\pi$ combinations (for the singly-tagged $D^-$
mesons). } \label{fig:dpkx}
\end{center}
\end{figure}

\begin{figure}[htbp]
\begin{center}
  \includegraphics[width=8cm,height=6cm]
{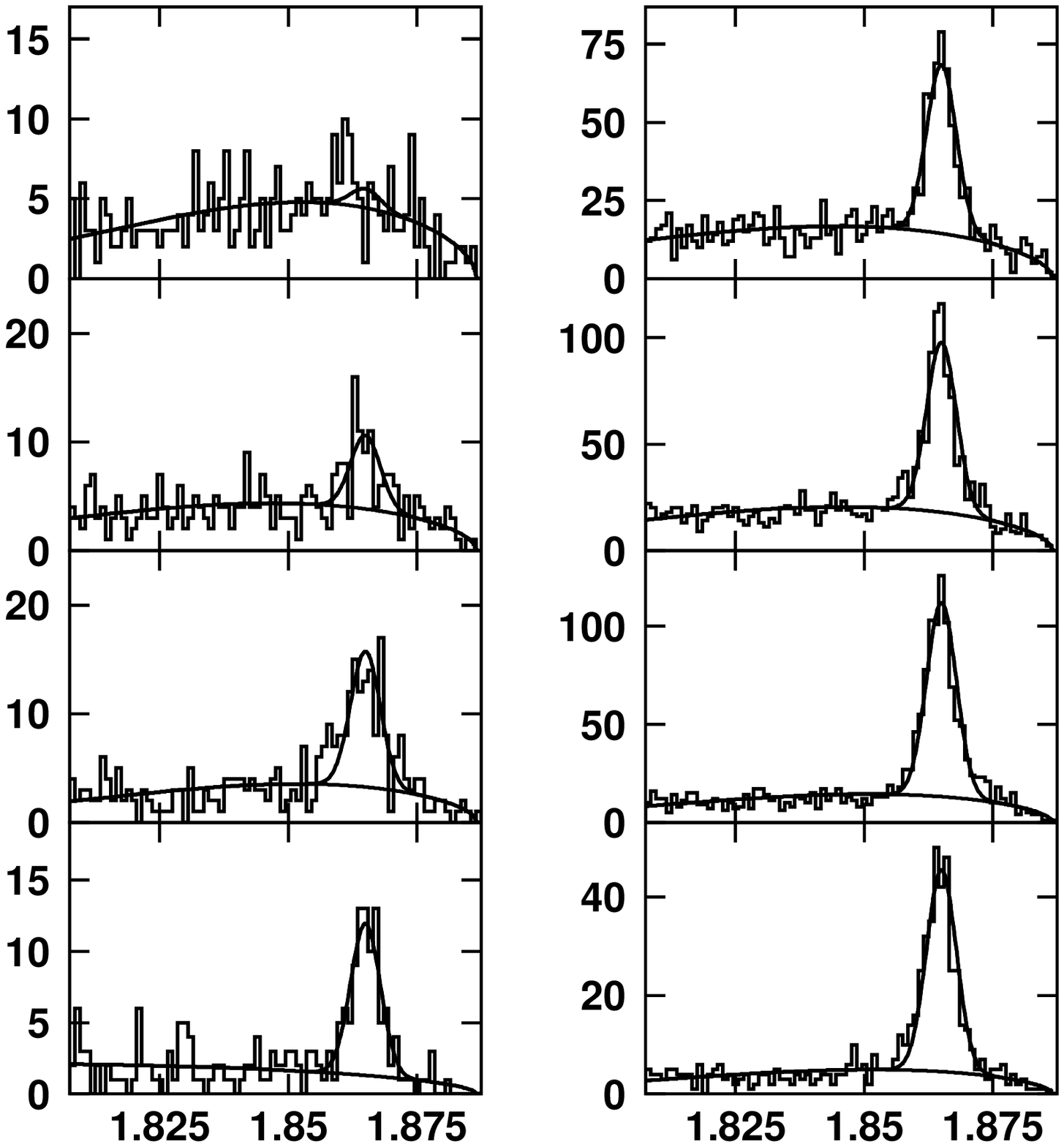}
      \put(-183,151){\small \bf a}
      \put(-183,114){\small \bf b}
      \put(-183,80){\small \bf c}
      \put(-183,43){\small \bf d}
      \put(-80,151){\small \bf a$^\prime$}
      \put(-80,114){\small \bf b$^\prime$}
      \put(-80,80){\small \bf c$^\prime$}
      \put(-80,43){\small \bf d$^\prime$}
      \put(-170,0){\bf \small Invariant Mass (GeV/$c^2$)}
      \put(-220,50){\rotatebox{90}{\bf \small Events/(0.001GeV/$c^2$)}}
\caption {\small The distributions of the invariant masses of the
$mKn\pi$ combinations for the events in which the candidates for the
right-sign (left) and wrong-sign (right) kaons with the momentum
within (a) 0.2-0.4 GeV/$c$, (b) 0.4-0.6 GeV/$c$, (c) 0.6-0.8 GeV/$c$
and (d) 0.8-1.0 GeV/$c$ ranges are observed in the system recoiling
against the $mKn\pi$ combinations (for the singly-tagged $\bar D^0$
mesons). } \label{fig:d0kx}
\end{center}
\end{figure}

\normalsize Figures \ref{fig:dpex}, \ref{fig:d0ex}, \ref{fig:dpkx}
and \ref{fig:d0kx} show the resulting distributions of the invariant
masses of the $mKn\pi$ combinations for the events in which the
candidates for the right-sign and wrong-sign electrons and kaons are
observed in the system recoiling against the $mKn\pi$ combinations.
Fitting to these mass spectra with a Gaussian function for the $\bar
D$ signal and a special function \cite{kpiev,dcrs,k0ev,kstev} for
the background, we obtain the numbers $N_{e,i}^{\rm obs}(R)$,
$N_{K,i}^{\rm obs}(R)$ of the right-sign electrons and kaons, and
the numbers $N_{e,i}^{\rm obs}(W)$, $N_{K,i}^{\rm obs}(W)$ of the
wrong-sign electrons and kaons observed from the data. In the fits,
the masses of the $D^-$ and $\bar D^0$ mesons are fixed at 1.8693
GeV/$c^2$ and 1.8650 GeV/$c^2$, and their mass resolutions are fixed
at 2.75 MeV/$c^2$ and 2.90 MeV/$c^2$, respectively, which are
determined from fitting the average $mKn\pi$ mass spectra. The
numbers $N_{\pi,i}^{\rm obs}(R)$ and $N_{\pi,i}^{\rm obs}(W)$ of the
right-sign and wrong-sign pions observed from the data are obtained
through a similar analysis.

The real yields $N_{e,i}^{\rm real}$, $N_{\pi,i}^{\rm real}$,
$N_{K,i}^{\rm real}$ of the right-sign or wrong-sign electrons,
pions and kaons are obtained through an unfolding procedure by the
matrix equation, \small
\[
\left(
\begin{array}{c}
N_{e,  i}^{\rm obs} \\
N_{\pi,i}^{\rm obs} \\
N_{K,  i}^{\rm obs} \\
\end{array} \right )
= \left(
\begin{array}{lcr}
\epsilon_{e,i} & f_{\pi \to e,i} & f_{K \to e,i} \\
f_{e \to \pi,i} & \epsilon_{\pi,i} & f_{K \to \pi,i}\\
f_{e \to K,i} & f_{\pi \to K,i} & \epsilon_{K,i} \\
\end{array} \right )
\left(
\begin{array}{c}
N_{e,  i}^{\rm real} \\
N_{\pi,i}^{\rm real} \\
N_{K,  i}^{\rm real} \\
\end{array} \right ).
\]
\normalsize Inserting $N_{a,i}^{\rm obs}(R)$, $\epsilon_{a,i}$ and
$f_{b \to a,i}$ in the matrix equation for the unfolding procedure,
we obtain the real yield $N_{e,i}^{\rm real}(R)$ of the right-sign
electrons in the $i$th momentum range. Similarly, using
$N_{a,i}^{\rm obs}(W)$, $\epsilon_{a,i}$ and $f_{b \to a,i}$, we
obtain the real yield $N_{e,i}^{\rm real}(W)$ of the wrong-sign
electrons in the $i$th momentum range. Subtracting $N_{e,i}^{\rm
real}(W)$ from $N_{e,i}^{\rm real}(R)$, yields the number $N^{\rm
real}_{D \to e^+X,i}$ of the inclusive semileptonic decay $D \to
e^+X$ in the $i$th momentum range. Adding the numbers of $N^{\rm
real}_{D \to e^+X,i}$ in all momentum ranges, we obtain the total
number $N^{\rm real}_{D \to e^+X}$ of the inclusive semileptonic
decay $D \to e^+X$, as summarized in Tab. \ref{tab:dex}.

\normalsize
\subsection{\bf Candidates for $D \to K^-X$ and $D\to K^+X$}
In fact, the wrong-sign and right-sign kaons observed in the system
recoiling against the singly-tagged $\bar D$ mesons correspond to
the decay $D \to K^-X$  and the decay $D \to K^+X$, respectively.
The above unfolding procedure can also give the real numbers
$N_{K,i}^{\rm real}(W)$ and $N_{K,i}^{\rm real}(R)$ of the
wrong-sign and right-sign kaons in the $i$th momentum range. Adding
these unfolded yields of the wrong-sign and right-sign kaons in all
momentum ranges yields the total numbers $N^{\rm real}_{D\to
K^{-,+}X}$ of the inclusive decays $D \to K^-X$ and $D \to K^+X$,
respectively. These numbers are summarized in Tabs. \ref{tab:dpkx}
and \ref{tab:d0kx}.

\small
\begin{table}[htbp]
\caption{The real numbers of the inclusive semileptonic decays
$D^+\to e^+X$ and $D^0\to e^+X$.}
\begin{center}
\small
\begin{tabular}{ c|c |c }  \hline
 $p_e$(GeV/$c$)      & $D^+ \to e^+X$ & $D^0 \to e^+X$ \\ \hline
 0.2$\sim$0.4  & $171.7 \pm 21.3$ & $ 97.9 \pm 24.0$ \\
 0.4$\sim$0.6  & $227.2 \pm 19.4$ & $131.0 \pm 17.2$ \\
 0.6$\sim$0.8  & $107.9 \pm 13.2$ & $ 57.4 \pm 11.4$ \\
 0.8$\sim$1.0  & $14.2 \pm 4.8$ & $  3.3 \pm 4.3$ \\ \hline
 Total         & $520.9 \pm 32.1$ & $289.6 \pm 31.9$ \\ \hline
\end{tabular}
\label{tab:dex}
\end{center}
\end{table}

\begin{table}[htbp]
\caption{The real numbers of the inclusive decays $D^+\to K^-X$ and
$D^+\to K^+X$.}
\begin{center}
\begin{tabular}{ c|c |c }  \hline
 $p_K$(GeV/$c$)   & $D^+ \to K^-X$ & $D^+ \to K^+X$ \\ \hline
 0.2$\sim$0.4 & $198.7 \pm 18.6$ & $ 44.4 \pm 11.9$ \\
 0.4$\sim$0.6 & $259.3 \pm 20.8$ & $ 44.7 \pm 13.8$ \\
 0.6$\sim$0.8 & $132.4 \pm 15.6$ & $ 72.9 \pm 15.4$ \\
 0.8$\sim$1.0 & $40.8 \pm 8.6$ & $ 26.7 \pm 12.8$ \\ \hline
 Total        & $631.2 \pm 33.1$ & $188.7 \pm 27.0$ \\ \hline
\end{tabular}
\label{tab:dpkx}
\end{center}
\end{table}

\begin{table}[htbp]
\caption{The real numbers of the inclusive decays $D^0\to K^-X$ and
$D^0\to K^+X$.}
\begin{center}
\begin{tabular}{ c|c |c }  \hline
 $p_K$(GeV/$c$)   & $D^0 \to K^-X$ & $D^0 \to K^+X$ \\ \hline
 0.2$\sim$0.4 & $406.5 \pm 27.2$ & $ 12.3 \pm 8.9$ \\
 0.4$\sim$0.6 & $603.6 \pm 32.4$ & $ 36.7 \pm 10.7$ \\
 0.6$\sim$0.8 & $758.9 \pm 34.3$ & $ 52.9 \pm 13.9$ \\
 0.8$\sim$1.0 & $328.9 \pm 23.6$ & $ 17.1 \pm 12.7$ \\ \hline
 Total        &$2097.9 \pm 59.3$ & $118.9 \pm 23.4$ \\ \hline
\end{tabular}
\label{tab:d0kx}
\end{center}
\end{table}

\normalsize
\section{\bf Results}
\subsection{\bf Monte Carlo efficiency}
 The efficiencies for reconstruction of the inclusive decays
$D^{+,0} \to K^{-,+} X$ and $D^{+,0} \to e^+ X$ are estimated by
Monte Carlo simulation. The Monte Carlo events are generated as
$e^+e^- \to D \bar D$, where $\bar D$ decays to the singly-tagged
$\bar D$ modes and $D$ decays to the $K^{-,+}(e^+) X$. The
efficiencies  are obtained by weighting the branching fractions
quoted from PDG06~\cite{pdg} and the numbers of singly-tagged $\bar
D$ events. These are $\epsilon_{D^+\to K^-X} = (48.1 \pm 0.9) \%$,
$\epsilon_{D^+\to K^+X} = (57.8 \pm 2.8) \%$, $\epsilon_{D^0\to
K^-X} = (51.6 \pm 0.4) \%$, $\epsilon_{D^0\to K^+X} = (48.4 \pm 2.4)
\%$, $\epsilon_{D^+\to e^+ X} = (64.4 \pm 1.0) \%$ and
$\epsilon_{D^0\to e^+ X} = (65.2 \pm 1.0) \%$,  where the error is
statistical.

\subsection{\bf Branching fractions}
 The branching fractions for the inclusive decay
$D^{+,0} \to K^{-,+}(e^+)X$ is determined by
\begin{eqnarray}
{BF(D^{+,0}} \to K^{-,+}(e^+)X)~~~~~~~~~~~~~~~~~~~~~~\nonumber\\
\ \ \ \ \ \  =\frac{N^{\rm real}_{D^{+,0} \to K^{-,+}(e^+)X}}
{N_{\bar D_{\rm tag}}\times\epsilon_{D^{+,0} \to K^{-,+}(e^+)X}},
\label{brinsemi}
\end{eqnarray}
where $N^{\rm real}_{D^{+,0} \to K^{-,+}(e^+)X}$ is the real number
of the inclusive decay $D^{+,0} \to K^{-,+}(e^+)X$, $N_{\bar D_{\rm
tag}}$ is the total number of the singly-tagged $D^-$ or $\bar D^0$
mesons, $\epsilon_{D^{+,0} \to K^{-,+}(e^+)X}$ is the detection
efficiency for $D^{+,0} \to K^{-,+}(e^+)X$. Inserting these numbers
in Eq. (\ref{brinsemi}), we obtain the branching fractions for these
inclusive decays to be
$$ BF(D^+\to K^-X) = (24.7 \pm 1.3 \pm 1.2) \%,$$
$$ BF(D^+\to K^+X) = (6.1 \pm 0.9 \pm 0.4)  \%,$$
$$ BF(D^0\to K^-X) = (57.8 \pm 1.6 \pm 3.2) \%,$$
$$ BF(D^0\to K^+X) = (3.5 \pm 0.7 \pm 0.3) \%,$$
$$ BF(D^+ \to e^+X) = (15.2 \pm 0.9 \pm 0.8) \%$$
and
$$ BF(D^0 \to e^+X) = (6.3 \pm 0.7 \pm 0.4) \% ,$$
where the first errors are statistical and the second systematic.
The systematic errors arise mainly from the uncertainties in the
tracking efficiency (2.0\% per track \cite{kpiev,dcrs,k0ev,kstev}),
the particle identification (0.5\% per track for kaon and pion ,
1.0\% per track for electron \cite{kpiev,dcrs,k0ev,kstev}), the
$\delta z$ selection criterion (1.8\% for the $D^+$ decays and 1.7
\% for the $D^0$ decays), the Monte Carlo statistics of the
detection efficiencies ($(0.8\sim5.0)\%$), the number of the
singly-tagged $\bar D$ mesons (3.0\% for $D^-$ and 4.5\% for $\bar
D^0$ \cite{kpiev,k0ev,kstev}) and the statistics of the
$\epsilon_{a,i}$ and $f_{b\to a,i}$ ($(1.5\sim3.8)\%$). In addition,
there is the uncertainty in the efficiency for identifying electron
and the probability for misidentifying electron as kaon or pion due
to the difference of event topologies between radiative Bhabha and
$D \bar D$ events. This uncertainty is estimated to be about 2.1\%
by analyzing radiative Bhabha and $D\bar D$ Monte Carlo events.
Adding these uncertainties in quadrature yields the total systematic
errors 4.9\%, 6.8\%, 5.5\%, 8.2\%, 5.1\% and 6.2\% for $D^+ \to K^-
X$, $D^+ \to K^+X$, $D^0 \to K^- X$, $D^0 \to K^+ X$, $D^+ \to e^+
X$ and $D^0 \to e^+ X$, respectively. The measured branching
fractions are well consistent with the measurements
\cite{MarkIII-ex,cleo-ex} from the MARKIII and CLEO Collaborations.

\subsection{\bf Ratios of $\frac{BF(D^+ \to e^+X)}{BF(D^0 \to
e^+X)}$ and $\frac{\Gamma(D^+ \to e^+X)}{\Gamma(D^0 \to e^+X)}$}

With the measured branching fractions for $D^+ \to e^+X$ and $D^0
\to e^+X$, we determine the ratio of the two branching fractions to
be
\begin{equation}
\frac{BF(D^+ \to e^+X)}{BF(D^0 \to e^+X)} = 2.41 \pm 0.30 \pm 0.18,
\end{equation}
where the first error is statistical and the second systematic
arising from some uncanceled uncertainties in the measurements of
the branching fractions.

Using the above ratio and the lifetimes of $D^+$ and $D^0$ mesons
\cite{pdg}, we obtain the ratio of their partial widths to be
\begin{equation}
\frac{\Gamma(D^+ \to e^+X)}{\Gamma(D^0 \to e^+X)} = 0.95 \pm 0.12
 \pm 0.07,
\end{equation}
where the first error is statistical and the second systematic. The
systematic error arises from the uncertainty in the measured ratio
of $BF(D^+ \to e^+X)/BF(D^0 \to e^+X)$, and the uncertainties in the
lifetimes of $D^+$ and $D^0$ mesons. This measured ratio is
consistent within error with those \cite{MarkIII-ex,cleo-ex}
measured by the MARKIII and CLEO Collaborations. This work and the
previous works reported in Refs. \cite{k0ev,k0muv} confirm that the
isospin conversation holds in the semileptonic decays of $D^+$ and
$D^0$ mesons.

\section{\bf Summary}
\vspace{-0.2cm}
 In summary, by analyzing the data of about 33 pb$^{-1}$
collected at and around 3.773 GeV with the BESII detector at the
BEPC collider, we have measured the branching fractions for the
inclusive $K^\pm$ decays of $D^+$ and $D^0$ mesons to be
 $BF(D^+\to K^-X) = (24.7 \pm 1.3 \pm 1.2)\%$,
 $BF(D^+\to K^+X) = (6.1 \pm 0.9 \pm 0.4) \%$,
 $BF(D^0\to K^-X) = (57.8 \pm 1.6 \pm 3.2) \%$
 and $BF(D^0\to K^+X) = (3.5 \pm 0.7 \pm
 0.3) \%$ with improved precision compared to those listed in PDG06
\cite{pdg}. We have also measured the branching fractions for the
inclusive semileptonic decays of $D^+$ and $D^0$ mesons to be
$BF(D^+ \to e^+X) = (15.2 \pm 0.9 \pm 0.8) \%$ and $BF(D^0 \to e^+X)
= (6.3 \pm 0.7 \pm 0.4) \%$, respectively. These yield the ratio of
their partial widths to be $\Gamma(D^+\to e^+X)/\Gamma(D^0 \to e^+X)
= 0.95 \pm 0.12\pm 0.07$, which supports the spectator model of the
$D$ meson decays.

\vspace{0.5cm}
\section{Acknowledgments}
The BES collaboration thanks the staff of BEPC for their hard
efforts. This work is supported in part by the National Natural
Science Foundation of China under contracts Nos. 10491300, 10225524,
10225525, 10425523, the Chinese Academy of Sciences under contract
No. KJ 95T-03, the 100 Talents Program of CAS under Contract Nos.
U-11, U-24, U-25, the Knowledge Innovation Project of CAS under
Contract Nos. U-602, U-34 (IHEP), the National Natural Science
Foundation of China under Contract  No. 10225522 (Tsinghua
University).

\end{document}